%% file: main.tex
\documentclass[sigconf,authorversion,nonacm]{acmart}

\AtBeginDocument{%
  \providecommand\BibTeX{{%
    \normalfont B\kern-0.5em{\scshape i\kern-0.25em b}\kern-0.8em\TeX}}}

\acmConference[Preprint]{}{September}{2025}
\acmBooktitle{Preprint} 

\input{preamble_lab}

\begin{document}

\title[Examining Interface Designs in Mobile Enhance Expressive Writing]{Examining Input Modalities and Visual Feedback Designs in Mobile Expressive Writing}

\settopmatter{authorsperrow=4}
\author{Shunpei Norihama}
\affiliation{%
  \institution{Interactive Intelligent Systems Laboratory}
  \institution{The University of Tokyo}
  \state{Tokyo}
  \country{Japan}
  \postcode{113-8656}
}
\email{norihama@iis-lab.org}

\author{Shixian Geng}
\affiliation{%
  \institution{Interactive Intelligent Systems Laboratory}
  \institution{The University of Tokyo}
  \state{Tokyo}
  \country{Japan}
  \postcode{113-8656}
}
\email{shixiangeng@iis-lab.org}

\author{Kakeru Miyazaki}
\affiliation{%
  \institution{Interactive Intelligent Systems Laboratory}
  \institution{The University of Tokyo}
  \state{Tokyo}
  \country{Japan}
  \postcode{113-8656}
}
\email{kakeru-miyazaki@iis-lab.org}

\author{Arissa J. Sato}
\affiliation{%
  \institution{Interactive Intelligent Systems Laboratory}
  \institution{The University of Tokyo}
  \state{Tokyo}
  \country{Japan}
  \postcode{113-8656}
}
\affiliation{
  \institution{University of Wisconsin-Madison}
  \city{Madison}
  \state{Wisconsin}
  \country{United States}
}
\email{asato@wisc.edu}

\author{Mari Hirano}
\affiliation{
  \institution{Ochanomizu University}
  \streetaddress{2-1-1 Ohtsuka}
  \state{Tokyo}
  \country{Japan}
  \postcode{112-8610}
}
\email{hirano.mari@ocha.ac.jp}

\author{Simo Hosio}
\affiliation{%
  \institution{University of Oulu}
  \streetaddress{Pentti Kaiteran katu 1}
  \city{Oulu}
  \country{Finland}
  \postcode{90570}
}
\affiliation{%
  \institution{Tokyo College}
  \institution{The University of Tokyo}
  \state{Tokyo}
  \country{Japan}
  \postcode{113-8656}
}
\email{simo.hosio@oulu.fi}

\author{Koji Yatani}
\affiliation{%
  \institution{Interactive Intelligent Systems Laboratory}
  \institution{The University of Tokyo}
  \state{Tokyo}
  \country{Japan}
  \postcode{113-8656}
}
\email{koji@iis-lab.org}

\renewcommand{\shortauthors}{Norihama et al.}

\begin{abstract}
Expressive writing is an established approach for stress management. Recently, information technologies, such as smartphones, have also been explored for expressive writing. 
Although mobile interfaces have the potential to support various daily writing activities, interface designs for mobile expressive writing and their effects on stress relief still lack empirical understanding. 
We examined the interface design of mobile expressive writing by investigating the influence of input modalities and visual feedback designs on usability and perceived cathartic effects through field studies. 
While our studies confirmed the stress-relieving effects of mobile expressive writing, our results offer important insights into interface design. 
We found keyboard-based text entry more suited and preferred over voice messages for its privacy and reflective nature.
Participants expressed different reasons for preferring different post-writing visual feedback depending on the cause and type of stress. 
This work advances interface design for mobile expressive writing and deepens understanding of its effects.

\end{abstract}



\maketitle

\section{Introduction}

Expressive writing is an established approach for mental well-being management, referring to textual disclosure of stressful experiences and their related emotions~\cite{pennebaker1997writing, pennebaker2007expressive}. 
It can provide people with a sense of stress relief~\cite{pennebaker2007expressive}, known as a \textit{cathartic effect}~\cite{breuer1955standard}.
Expressive writing originated from pen and paper~\cite{pennebaker2007expressive}.
But the practice has expanded to information technologies~\cite{hardey2002story}, including mobile apps~\cite{wang2018mirroru, wu2018soften, macisaac2022writing} and chatbots~\cite{DigitalWorkplace, park2021wrote} as mental well-being is becoming an important issue to the general user populations.
Digital implementations maintain many of the traditional benefits of expressive writing while adding advantages of social sharing, such as engagement, reflection, and guidance~\cite{park2021wrote}.
While digital implementations of expressive writing have demonstrated cathartic effects~\cite{DigitalWorkplace}, there still exists a research gap to fully leverage the capability of recent information technology.   
In particular, as smartphones are now highly ubiquitous, optimizing their use can offer an opportunity for people to perform expressive writing at hand, promoting routine stress management and contributing to their mental well-being.

Despite its promising potential, there is a lack of understanding in mobile interface designs for expressive writing.
Regarding the input modalities, existing research has revealed advantages of speech-based input for mobile interfaces in speed~\cite{ruan2018comparing}, workload~\cite{rzepka2022voice}, and content richness~\cite{luo2022promoting}.
However, it remains underexplored how such an input modality could contribute to user experience in mobile expressive writing.
Prior work also shows that verbal expression may provoke a sense of releasing stress and enhance cognitive change~\cite{esterling1994emotional, murray1989emotional}, but it is unknown how speech-based expressive writing compares to traditional text-based input with respect to cathartic effects and user experience.
Feedback to users after disclosure is another key design component for mobile expressive writing.
Clinical psychology has shown that deleting objects tied to emotion helps to resolve negative moods~\cite{Brinol2013, lee2011wiping, soesilo2021no}.
People may write a note with their negative emotion and put it into a shredder or simply tear it up by hand~\cite{nussbaum1995objectification, soesilo2021no}, which can provide an explicit feeling of removing or destroying the negative experience~\cite{Brinol2013, lee2011wiping}.
Translating this mechanism into mobile interfaces presents a design challenge: balancing the emotional impact with the level of user effort required.
To address this, we introduced the concept of \textit{passive intervention}~\cite{zhao2023affective} and compared it with interactive visual feedback and a condition offering only textual responses.
Our exploration
can inform future design strategies for technology-supported emotional regulation.

This paper aims to better understand the cathartic effect of expressive writing on mobile devices.
More specifically, we design our studies to answer the following two questions:

\begin{itemize}
    \item[RQ1.] How do speech and text-based input in mobile expressive writing influence the stress-relieving effect, the content of expression, and the perceived usability and preference?
    \item[RQ2.] How do different visual feedback designs in mobile expressive writing influence the stress-relieving effect, and the perceived usability and preference?
\end{itemize}

To answer these questions, we conducted two field studies (15 and 23 days long with 28 and 36 participants, respectively).
Our first study examined the effect of keyboard-based and speech-based input, and the second experiment focused on the effect of post-disclosure interaction. 
Our studies highlighted the unique benefits of each of the input modalities and post-disclosure interaction design instead of the superiority of one of those.
Our results suggest future research directions in mobile expressive writing.
The contributions of this paper, therefore, are two-folded: 

\begin{itemize}
    \item Empirical evidence on how input modalities and post-disclosure visual feedback designs influence the cathartic effect and user experience in mobile expressive writing, along with a discussion on underlying psychological processes.
    \item Design and research implications based on our two studies, highlighting the necessity of considering not only stress levels but also stress types in the design of
    technology-supported expressive writing and stress coping.
\end{itemize}

\section{Related Work}\label{sec:related}

\subsection{Expressive Writing: Outcomes, Mechanisms, and Protocols}\label{subsec:rw-stresscoping}

Writing can have a therapeutic effect under certain conditions. The therapeutic use of writing, known as \textit{expressive writing}~\cite{pennebaker1997writing, pennebaker2007expressive}, is also referred to as \textit{written emotional disclosure}~\cite{frisina2004meta} or \textit{therapeutic writing}~\cite{wright2001mastery}. Pennebaker et al. introduced expressive writing as a therapeutic method for PTSD and found it improved physical health~\cite{pennebaker1986confronting}. Since then, research has revealed various benefits, including reduced anxiety~\cite{reynolds2000emotional} and depression~\cite{soliday2004expressive}, cognitive and self-esteem changes~\cite{pennebaker1997writing, murray1994emotional}, decreased aggression~\cite{kliewer2011school}, and improved physical conditions~\cite{pennebaker1986confronting}. However, while expressive writing promotes well-being, it may also initially increase distress and negative mood~\cite{esterling1999empirical}, though studies suggest a subsequent reduction in depressive symptoms~\cite{reinhold2018effects}. These complex and sometimes contradictory outcomes have led to extensive discussions on the underlying processes of expressive writing.

Three common theories attempt to explain the effects of expressive writing: \textit{inhibition theory}, \textit{exposure theory}, and \textit{cognitive processing}~\cite{reinhold2018effects}. Early research followed \textit{inhibition theory}, suggesting that suppressing thoughts and emotions increases mental workload, and therefore, expressing them improves physical and mental health~\cite{pennebaker1986confronting}. \textit{Exposure theory} posits that expressive writing functions as imaginative exposure to negative emotional events, helping individuals become desensitized. Through this process, people unlearn past stress responses and adopt new ones, potentially explaining expressive writing’s effects~\cite{sloan2004taking}. \textit{Cognitive processing} focuses on content, arguing that constructing a coherent narrative helps individuals organize experiences, shift perspectives, and gain clarity~\cite{pennebaker2007expressive, niles2016writing}. These theories suggest that expressive writing encourages confronting and releasing suppressed emotions. While it may temporarily intensify negative moods, it ultimately helps restructure cognition around stressful events, yielding various benefits.

Researchers have also explored the boundary conditions of expressive writing to deepen understanding. In Pennebaker's basic paradigm, individuals write about their deepest thoughts and feelings regarding traumatic or stressful experiences~\cite{pennebaker2007expressive}. Participants are assured confidentiality and encouraged to focus on emotions rather than spelling, structure, or grammar. Each session lasts 15–30 minutes, repeated over 3–5 days. Building on this protocol, research has examined various influencing factors~\cite{pennebaker2007expressive, smyth2008exploring}, expanding the scope of expressive writing. Notably, it is not limited to traumatic or negative experiences but can include positive events~\cite{burton2004health, king2000writing} or others' traumatic episodes~\cite{greenberg1996health}. Session duration also varies; even two-minute writing sessions over two days have shown health benefits~\cite{burton2008effects}. Additionally, session frequency is flexible, with a single 10–15 minute session with positive framing or a one-hour intensive session yielding similar benefits to the traditional multi-day approach~\cite{chung2008variations, smyth2008exploring}.

\subsection{Technology-Supported Expressive Writing}\label{subsec:rw-techew}

Research has explored the use of information technology to support expressive writing~\cite{hardey2002story, sauter2014s}. Experiments with \textit{EmotionDiary}, an expressive writing application for online social networking services (SNS), revealed that it was particularly effective for individuals with prior experience in emotional disclosure on SNS~\cite{lee2016insights}. Another study designed chatbot interactions for expressive writing, showing that a responsive chatbot enhanced emotional disclosure more than writing on a document~\cite{park2021wrote}. Other applications support journaling~\cite{wang2018mirroru}, reflection~\cite{mcduff2012affectaura}, and recollection~\cite{peesapati2010pensieve} of both everyday emotional experiences and trauma-related stress.
These technologies can maintain the benefits associated with traditional expressive writing while providing potential advantages typically afforded by social sharing, such as enhanced engagement, reflection, and guidance, without incurring the risks often associated with human-based social disclosure, including stigma and loss of privacy~\cite{park2021wrote}.

Research also compared different chatbot interventions for workplace stress: taking breaks, promoting mindfulness, and emotional expression~\cite{DigitalWorkplace}. Though emotional expression was most effective in reducing stress, it was rated lowest due to the high effort required and increased stress during reflection. Technology-supported expressive writing, revealed emotions and reduced stress, but caused short-term emotional decline and was burdensome. Thus, there is still a need to improve mobile expressive writing interfaces for cathartic benefits.

\subsubsection{The Effect of Input Modalities in  Expressive Writing}

\label{subsec:rw-input}
Verbal forms of expressive writing, so-called \textit{expressive talking}~\cite{harrist2007benefits, langer2012expressive}, yield unique outcomes. Previous research indicates verbal expressions (e.g., speaking into a tape recorder) may enhance cognitive change and immune function more effectively than written expressions (e.g., writing on paper)~\cite{esterling1994emotional}, or at least offer comparable health benefits~\cite{murray1994emotional, lyubomirsky2006costs}. The expression mode also influences word choice: Talking involves more words~\cite{murray1994emotional}, fewer negative emotional words~\cite{esterling1994emotional}, and a more coherent narrative~\cite{newton2015emotional} than writing. In expressive writing, overuse of negative emotional words leads to negative results, while creating a coherent story offers advantages~\cite{pennebaker2007expressive}; these linguistic features may be linked to unique outcomes.

Many expressive writing technologies use physical or virtual keyboards~\cite{abd2019overview, park2021wrote, DigitalWorkplace}. Brewin and Lennard found that typing, compared to pen and paper, led to fewer negative effects but also less self-disclosure~\cite{brewin1999effects}. This highlights the need to explore the impact of different input methods. Speech-based input is promising on smartphones~\cite{perficient2020voiceusage}, outperforming keyboards for short messages~\cite{ruan2018comparing}, aiding search with conversational agents~\cite{rzepka2022voice}, and assisting in food journaling~\cite{luo2022promoting}.
Research on verbal disclosure versus written self-disclosure suggests essential differences.
For example, Iacovelli et al. observed stronger physiological benefits from face-to-face disclosure compared to text messaging~\cite{iacovelli2012disclosure}.
Similarly, comparisons between therapy interviews (with therapists) and expressive writing (self-administered) indicate that interviews promote positive affect~\cite{donnelly1991cognitive, murray1989emotional} and cognitive change~\cite{murray1989emotional}. 
These projects indicate the potential for further exploration of alternative input modalities, particularly speech, within the context of mobile expressive writing.

\subsubsection{Feedback Effects in Expressive Writing and Materializing Emotional Catharsis Method}\label{subsec:rw-feedback}

Researchers have studied feedback's role in enhancing expressive writing, which is inherently self-directed~\cite{pennebaker1986confronting, pennebaker2007expressive}. Feedback has been introduced post-writing, especially for learners with disabilities~\cite{gersten2001teaching, macarthur1996integration}, using input from teachers~\cite{sawyer1992direct} or peers~\cite{wong1996teaching, o2018suddenly}.
AI-driven conversational feedback, like Park et al.'s Responsive Chat, encourages self-awareness and reflection by posing questions on emotions and impacts~\cite{park2021wrote}. However, user privacy is a major issue in adaptive feedback~\cite{pennebaker2007expressive}.
While technology that offers a low-cost, privacy-preserving feedback solution exists~\cite{bond2012automated, park2021wrote}, feedback designs that do not engage in the content of disclosure would liberate developers and users from privacy concerns.

Psychological research highlights the cathartic effect of materializing emotions as tangible entities and symbolically destroying them. Prior work has shown that additional actions on expressed negative emotions, such as physically discarding them, can enhance stress relief~\cite{Brinol2013, lee2011wiping, soesilo2021no, kanaya2024anger}.
Bri{\~n}ol et al.\cite{Brinol2013} found that even digital interactions, such as dragging files representing negative emotions to the trash, had a greater effect than merely imagining their disposal. Digital technology further extends this concept by enabling flexible expressive writing and interactive emotional processing. In virtual reality, physically punching objects with stressful messages promoted relaxation\cite{grieger2021trash}, while Bj{\"o}rling et al.\cite{bjorling2019thought} developed a VR system where teenagers burned negative thoughts as leaves to reduce stress. However, most mobile device studies focus on limited feedback\cite{park2021wrote, DigitalWorkplace}, and the impact of visual feedback after emotional sharing remains underexplored.

There has been ongoing debate regarding the key factors that contribute to the effectiveness of materialized emotional disposal.
Lee et al.~\cite{lee2021grounded} reported that physical engagement enhanced cathartic effects.
According to grounded cognition theory~\cite{barsalou2008grounded}, intense physical movements reinforce the perceived meaning of the action, facilitating a deeper sense of \textit{de-fusion} (i.e., decouple) from negative emotions~\cite{hatvany2018becoming}.
From this perspective, interactive interventions that promote a sense of agency~\cite{moore2016sense} on disposal may be especially effective in mobile interface design.
In contrast, Bri{\~n}ol et al.~\cite{Brinol2013} found that the interpretation of disposal may play a more critical role than the sensorimotor experience itself.
According to this view, simply presenting the image of emotional disposal could be sufficient, even when the system carries out the action rather than the user.
However, the psychological effect of such system-initiated disposal remains underexplored~\cite{lee2011wiping}.
Additionally, prior research suggests that \textit{passive} interventions, which do not require active user engagement, can be effective in high-stress contexts~\cite{zhao2023affective}.
Given that minimal-effort interventions are preferable in everyday stress management applications~\cite{DigitalWorkplace}, it is vital to investigate how system-driven, passive interventions differ in psychological impact from interactive ones.

\section{Study 1: Examination of Input Modalities}\label{sec:study1}

As discussed in the previous section, while voice input may facilitate emotional disclosure and cognitive processing, prior studies have mainly focused on keyboard input, leaving its effects unexplored.
To address this gap and answer RQ1, we conducted a user study investigating how different input modalities affect the cathartic effect and usability in expressive writing.
This section introduces the design of our first user study.
The following study protocol was approved by the Institutional Review Board of the first author's institution.

\subsection{System Implementation}\label{subsec:study1systemdesign}

\begin{figure*}[tbp]
    \centering
    \includegraphics[width=0.65\linewidth]{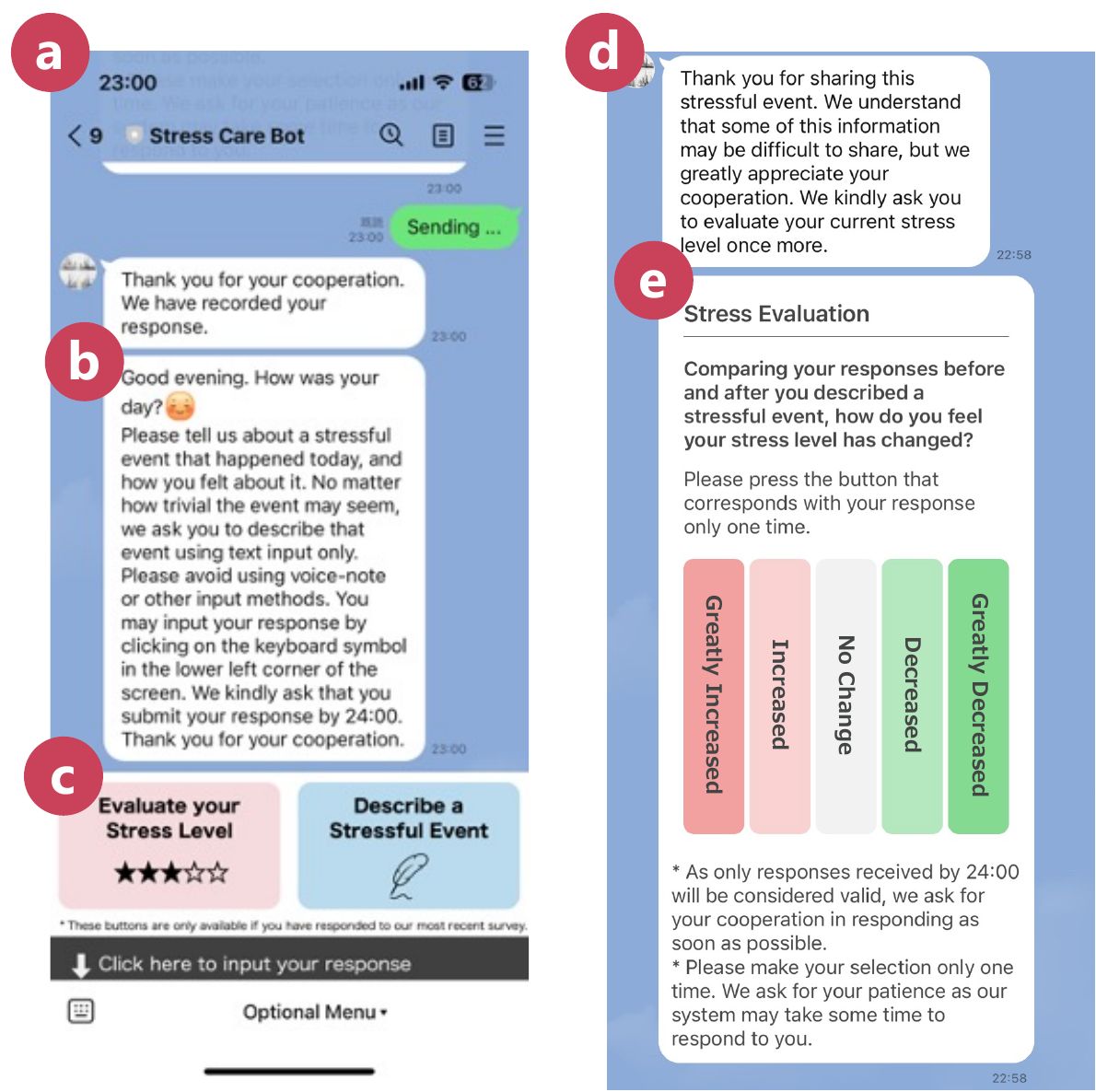}
    \caption{The screenshots of our system:
(a) The original interfaces were in Japanese, and we translated the text into English for this report.
(b) Messages prompting participants to share what made them stressed and how they felt.
(c) A menu at the bottom of the screen allowed participants to report their stress level or express stressful events and feelings at any time.
(d) After expressing their emotions, participants received two messages: the first acknowledged and appreciated their sharing.
(e) The second asked them to evaluate how their stress level had changed before and after their expression.
    }
    \Description[Two screenshots of our interface we used in Study 1.]{Two screenshots of our interface we used in Study 1. The first screenshot with areas labeled (a) to (c), and another screenshot with areas labeled (d) and (e). Section (a) dispalys the general chat UI of the system, showing the message sent from the system on the left side and message users sent on the right side. Section (b) displays the message sent by the system saying: "Good evening. How was your day? Please tell us about a stressful event that happened today, and how you felt about it. No matter how trivial the event may seem, we ask you to describe that event using text input only. Please avoid using voice-note or other input methods. You may input your response by clicking on the keyboard symbol in the lower-left corner of the screen. We kindly ask that you submit your response by 24:00. Thank you for your cooperation." Section (c) displays a menu at the bottom of the screen with two buttons: "Evaluate your Stress Level" with a five-star icon, and "Describe a Stressful Event" with a pen and paper icon. Below these buttons is text saying, "* These buttons are only available if you have responded to our most recent survey." and a clickable link saying, "Click here to input your response", followed by a "Optional Menu" button. Section (d) shows a reply message to the expressive writing that says, "Thank you for sharing this stressful event. We understand that some of this information may be difficult to share, but we greatly appreciate your cooperation. We kindly ask you to evaluate your current stress level once more." Section (e) shows a message for "Stress Evaluation" with buttons on it. The message starts with "Comparing your responses before and after you described a stressful event, how do you feel your stress level has changed? Please press the button that corresponds with your response only one time." Below this text are five buttons labeled "Greatly Increased" in red, "Increased" in light red, "No Change" in gray, "Decreased" in light green, and "Greatly Decreased" in green. A note at the bottom reads, "* As only responses received by 24:00 will be considered valid, we ask for your cooperation in responding as soon as possible. * Please make your selection only one time. We ask for your patience as our system may take some time to respond to you."}
    \label{fig:UI}
\end{figure*}

We developed a smartphone-based system for expressive writing (Figure~\ref{fig:UI}). We chose text chat interaction over document writing to enhance emotional disclosure~\cite{park2021wrote}. LINE was selected as the platform due to its popularity in Japan, where our study was conducted. Participants shared daily stress experiences and feelings using a specific input method (Figure~\ref{fig:UI}b). After completing their writing, they rated perceived stress changes, indicating the momentary cathartic effect (Figure~\ref{fig:UI}e).
The system also sent reminders.
We predefined all messages, using emojis to increase credibility~\cite{beattie2020bot} and enjoyment~\cite{fadhil2018effect}.

\subsection{Input Modality Conditions}\label{subsec:inputmodality}
Participants used a soft keyboard to type in the \textit{Keyboard} condition and sent voice messages in the \textit{Voice} condition. Conducted in Japan, the study allowed any keyboard design for realistic use, including non-QWERTY options like flick input. For voice messages, they used LINE's existing functionality. Participants were prohibited from using speech recognition for text input since LINE could not distinguish between speech-recognized and keyboard-entered text. Whisper\footnote{We decided to use Whisper because it exhibited a high level of recognition accuracy (approximately 94\%) for Japanese at the time of our analysis.}~\cite{radford2022robust} was used for transcribing voice data.

\subsection{Experiment Design}\label{subsec:study1-design}

\subsubsection{Independent Variables}\label{subsubsec:study1-design-iv}
We established two input conditions for the study: using soft keyboards (\textit{Keyboard}) and voice message (\textit{Voice}). These methods are key text entry approaches in computers and smart devices~\cite{yu2019almost}. While soft keyboards dominate smartphone text entry, voice input is gaining popularity~\cite{perficient2020voiceusage}. They also mirror communication styles in therapy, such as writing about stress or speaking with a therapist.

\subsubsection{Dependent Variables for Stress-relieving Effect and Content}\label{subsubsec:study1-design-dv2}

As it is known that expressive writing leads to immediate negative mood and eventual stress reduction~\cite{esterling1994emotional, pennebaker2007expressive}, we captured stress reduction effects immediately after the intervention and in the post-experiment questionnaire.

\paragraph{Immediate Stress Change}
To assess the cathartic effect of expressive writing, participants answered a question about changes in stress levels before and after describing a stressful event (Figure~\ref{fig:UI}e), based on previous stress coping research~\cite{tong2023just}. Responses were on a five-point Likert scale from -2 (greatly decreased) to 2 (greatly increased), completed after each writing task.

\paragraph{Post-experiment Cathartic Effect}
We assessed the cathartic effect post-experiment by asking, based on Howe~\cite{DigitalWorkplace}, if ``\{\textit{Keyboard/Voice}\} input was helpful for reducing my stress,'' using a five-point Likert scale (1: strongly disagree to 5: strongly agree) in the questionnaire.

\vspace{0.5\baselineskip}
We collected two parameters identified as meaningful for health outcomes~\cite{esterling1994emotional, murray1994emotional}, enabling us to understand how participants expressed and alleviated stress and emotions across different conditions.

\paragraph{Length}
We determined the length of each expressive writing by counting characters in the original Japanese text, with one English word equating to 2--2.5 Japanese characters.

\paragraph{Emotional Words}
To assess participants' emotional disclosure in writing, we analyzed positive and negative emotional word usage with the Linguistic Inquiry and Word Count (LIWC) program \cite{pennebaker2001linguistic, igarashi2022jliwc}. For each participant and input modality, descriptions were word-segmented using MeCab \cite{kudo2005mecab} and processed with LIWC. We measured the percentage of positive/negative emotion words according to LIWC results.

\subsubsection{Dependent Variables for Usability and Preference}\label{subsubsec:study1-design-dv1}
We gathered three quantitative measures in the post-experiment questionnaire to assess participants' perceived usability and preference for each input modality.

\paragraph{Ease of Use}
To assess ease of use, we used the question ``It was easy to enter stress by \{\textit{Keyboard/Voice}\},'' aligning with earlier usability studies of digital coping methods~\cite{DigitalWorkplace}. Responses were on a five-point Likert scale (1: ``It was difficult to do'' to 5: ``It was easy to do.'').

\paragraph{Mental Workload}
We used NASA-TLX to capture the perceived mental workload associated with each input modality condition~\cite{hart1988development}.

\paragraph{Preference}
To assess input modality preferences, we asked: ``Which input method would you prefer, \textit{Keyboard} or \textit{Voice}?''

\subsection{Procedure}

On the first day, participants first signed the consent form, received a withdrawal of consent form, and were informed that they could discontinue the study at any time, with any collected data discarded, and would still be compensated for any completed tasks.
Then they completed a pre-experiment questionnaire (see Appendix~\ref{appendix:study1-questionnaire}), and confirmed their ability to use our system.
The questionnaire included the Perceived Stress Scale (PSS)~\cite{cohen1994perceived} to assess initial stress levels. We explained that the study's primary goal was to collect stress-related emotions, avoiding bias about input modality effects. Afterward, we provided a link and QR code to access to our system.
We split participants into two balanced groups by gender, age, and occupation to counterbalance input modality presentations.

From Day 2 to Day 14, participants shared stressful experiences and emotions, but only submissions using the assigned input modality were accepted.
We asked questions daily between 6 and 7 p.m., with reminders sent to incomplete participants from 9 to 10 p.m. To reduce weekday stress effects~\cite{brantley1988daily}, participants alternated weekly conditions.
Participants who failed to respond received follow-up messages through the crowdsourcing platform.
On Day 8, input modalities were switched. Conditions lasted 7 and 6 days, aligning the start and end on weekends to facilitate easier participation.

On the final day, we revealed the study's true objectives in a debriefing. Participants were assured of full compensation even if they chose to withdraw, yet none did.
They signed a new consent form and completed a post-experiment questionnaire (Table \ref{tab:study1-questionnaire} in Appendix \ref{appendix:study1-questionnaire}), covering \textit{ease of use}, \textit{mental workload}, \textit{post-experiment cathartic effect} for each input modality, and \textit{preference} on input modalities, along with PSS questions. Qualitative data were also gathered through open-ended questions about their thoughts on input conditions and preferences.
These data aimed to provide deeper insights into the quantitative results while maintaining participant anonymity and encouraging candid, reflective responses.

We collected the consent forms and questionnaires through Google Forms and communicated with participants via the crowdsourcing platform.
We offered all participants approximately 19 USD in the local currency for the completion of the study.

\subsection{Participant recruitment}

We recruited participants through Crowdworks, one of the most widely used crowdsourcing platforms in Japan.
Majima et al.~\cite{majima2017conducting} confirmed that the subjects available through Crowdworks are sufficiently reliable for behavioral research.
We conducted no screening based on participants’ mental health history, as the study targeted a broad general population.
We recruited 28 participants (14 female and 14 male; PA1 -- 28) in their 20s to 60s (Table \ref{tab:demographics1}).

\begin{table*}[t]
    \centering
    \caption{Demographics information of participants in Study 1.}
    \label{tab:demographics1}
    \begin{tabular}{|c||c|c|}
    \hline
    Groups&Gender&Age Ranges \\ \hline\hline
    Group 1 (\textit{Keyboard} $\rightarrow$ \textit{Voice})&
    F: 7, M: 7 &
    20s: 2, 30s: 4, 40s: 3, 50s: 3, 60s: 2\\ \hline
    Group 2 (\textit{Voice} $\rightarrow$ \textit{Keyboard})&
    F: 7, M: 7 &
    20s: 2, 30s: 4, 40s: 4, 50s: 3, 60s: 1\\ \hline
    \end{tabular}
\end{table*}

\subsection{Ethical Considerations}

This study was approved by the Institutional Review Board of the first author’s institution, and all procedures complied with ethical guidelines for research involving human participants.

To avoid introducing potential biases, we employed a deception study protocol.
At the beginning of the study, we explained to our participants that the objective of the study was to collect stress-related emotional expressions.
At the end of the study, we disclosed the true objective of the study (i.e., a comparison of input modalities), and offered them an explicit opportunity for withdrawal without any negative consequence to them (e.g., full compensation would be offered even if they decided to withdraw).
We then asked them to consent to the disclosed experimental protocol and data use.
All the participants consented after this debriefing.

Before participation and after debriefing, all individuals provided informed consent through an online form.
Participants were informed that they could withdraw from the study at any time without penalty.
In the event of withdrawal, any data already collected would be discarded.
A separate withdrawal-of-consent form was also provided, and participants were compensated for all tasks completed prior to withdrawal.

We did not collect any personally identifiable information.
Participants were identified only by pseudonymous usernames assigned by the crowdsourcing platform.
These were used solely for managing participation and compensation.
All data were securely stored and used exclusively for this study.

To minimize potential psychological distress, we informed participants that they could stop the experiment at any time if they felt discomfort, and that we would also terminate participation if signs of excessive stress were observed.
Participants were compensated fairly based on the study duration and workload.

\subsection{Data Analysis}

To compare quantitative data collected in each condition or at pre- and post-experiment, we used non-parametric statistical tests.
For analyzing qualitative data, two of the authors performed a thematic analysis~\cite{braun2006using}.
The first author read through the free-text responses from the post-experiment questionnaire and developed a codebook with respect to factors that determined participants' preferences.
Then, the first author and another author reviewed the codes and discussed until a consensus was reached.  
Finally, we categorized the codes into themes.

\section{Study 1 Results}

In this section, we present the results of the first user study described in the previous section, focusing on usability and cathartic effects.
We collected 483 expressive writing entries and 479 data points on immediate stress change, i.e. a response rate of 98.9\%.
Among 483 expressive writing entries, 244 and 239 were entered under the \textit{Keyboard} and \textit{Voice} conditions, respectively.
The average scores for PSS before and after the experiment were 28.5 ($SD$ = 7.87) and 27.6 ($SD$ = 8.95), respectively, and our Wilcoxon signed-rank test did not find a significant difference ($W$ = 1, $Z$ = 0.250, $p$ = .27, $r$ = 0.0473).
The total length of responses across four open-ended questions averaged approximately 306 characters per participant, ranging from 36 to 799 characters, with a response rate of 100\%.

\subsection{Perceived Usability \& Preference}\label{subsubsec:study1-results-usability}
The \textit{Keyboard} condition showed better usability than the \textit{Voice} condition and thus was preferred.
We also extracted the rationale for their preference by analyzing the qualitative data.

\begin{figure*}[tbp]
    \centering
    \begin{minipage}[t]{0.32\linewidth}
      \centering
      \includegraphics[keepaspectratio, width = 0.77\linewidth]{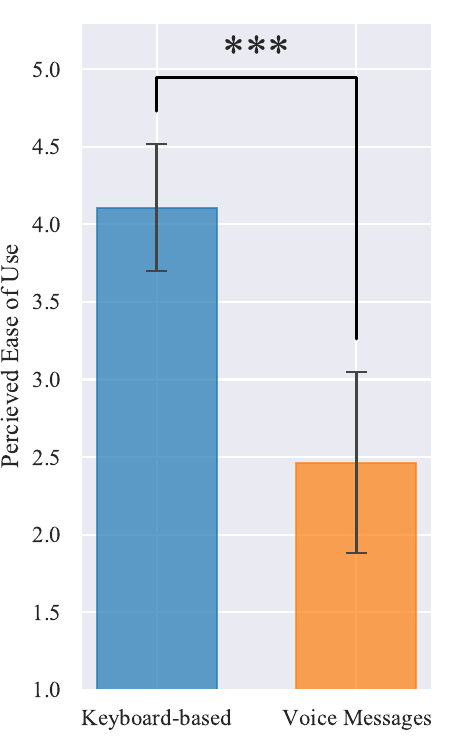}
      \subcaption{The average ease of use scores.}
      \label{fig:exit-easiness}
    \end{minipage}
    \hspace{0.1cm}
    \begin{minipage}[t]{0.32\linewidth}
      \centering
      \includegraphics[keepaspectratio, width = 0.77\linewidth]{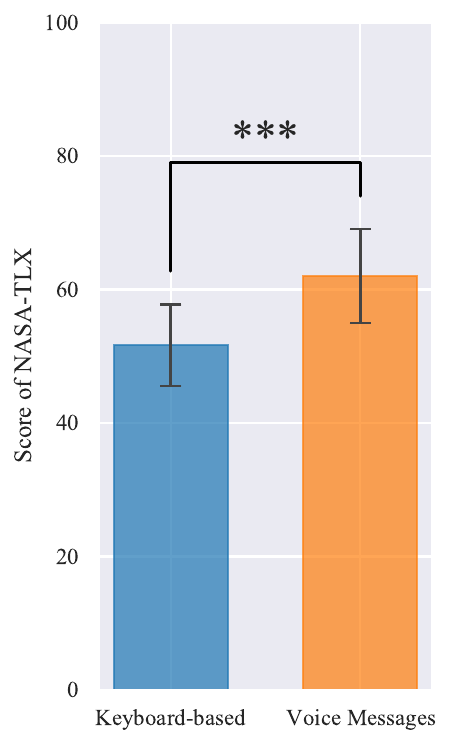}
      \subcaption{The average mental workload scores.}
      \label{fig:exit-nasatlx}
    \end{minipage}
    \hspace{0.1cm}
    \begin{minipage}[t]{0.32\linewidth}
      \centering
      \includegraphics[keepaspectratio, width = 0.77\linewidth]{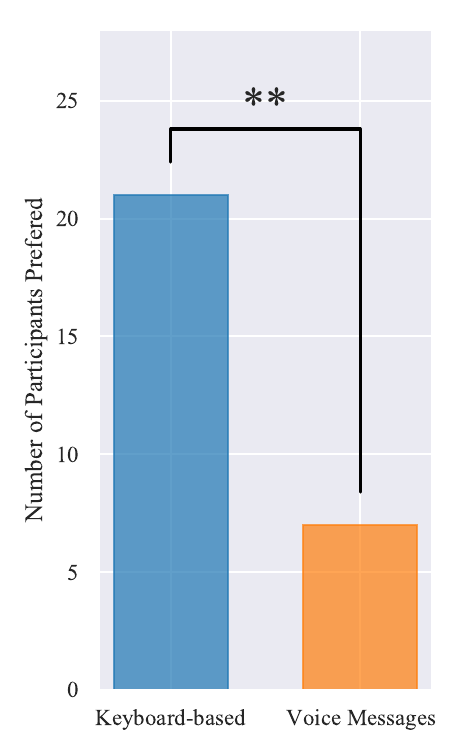}
    \subcaption{The preference votes on the two input modalities.}
    \label{fig:exit-preferable}
    \end{minipage}
  \caption{Results related to usability and preferences of the two input modalities. 
  All were collected through the post-experiment questionnaire.
  All error bars indicate 95\% confidence interval. (*: $p < .05$, **: $p < .01$, ***: $p < .001$)}
  \label{fig:exitsurvey}
\end{figure*}

\paragraph{Ease of Use}
Participants felt keyboard-based text entry was easier to use than sending \textit{Voice} messages.
The average values of \textit{ease of use} were 4.11 ($SD$ = 1.10) and 2.46 ($SD$ = 1.57) for the \textit{Keyboard} and \textit{Voice} conditions, respectively (Figure~\ref{fig:exit-easiness}).
Our Wilcoxon signed-rank test found a significant difference ($W$ = 1, $Z$ = -4.18, $p$ < .001, $r$ = 0.79).

\paragraph{Mental Workload}
In line with the ease of use, keyboard-based text entry required less mental workload.
The average values of \textit{mental workload} were 51.7 ($SD$ = 16.5) and 62.0 ($SD$ = 19.0) for the \textit{Keyboard} and \textit{Voice} conditions, respectively (Figure \ref{fig:exit-nasatlx}).
Our Wilcoxon signed-rank test found a significant difference ($W$ = 1, $Z$ = -3.46, $p$ < .001, $r$ = 0.65).
There were significant negative correlations between the ease of use and the mental workload scores in both the \textit{Keyboard} ($Spearman's\:r(26)$ = -0.39, $p$ < .05) and \textit{Voice} ($Spearman's\:r(26)$ = -0.43, $p$ < .05) conditions.

\paragraph{Preference}\label{subsubsec:study1-result-preference}
Twenty-one participants preferred keyboard-based text entry to voice messages while seven preferred voice messages (Figure~\ref{fig:exit-preferable}). 
Our chi-square test revealed that participants preferred the \textit{Keyboard} condition significantly more compared to the \textit{Voice} condition ($\chi^2(1)$ = 7.0, $p$ < .01).

\vspace{0.5\baselineskip}
We performed qualitative analysis on the comments we obtained from our participants in the post-experiment questionnaire.
We identified three major themes that highlighted the two important factors in the \textit{Keyboard} text entry method: \textit{privacy friendliness}, and \textit{reflection through typing}.

Participants found voice message input unsuitable as it often involved private or sensitive emotional disclosures, making them hesitant to use it in public spaces, especially when expressing negative emotions about stressful events at work or in relationships, as PA1 mentioned: ``\textit{I found it very challenging to use voice input because I didn't want others to hear me, so I had to wait for moments when there were no people around. This was especially difficult when I was at work in the company.}''

Typing with keyboards enabled participants to refine their thoughts and feelings by reviewing and modifying content objectively on their screens, as PA12 commented: ``\textit{I was able to write about the events of the day while recalling them in detail, which allowed me to organize my emotions.}''

Some participants said organizing the content while speaking was challenging, and they sometimes found themselves not knowing what to say in the middle of recording.
PA1 commented: ``\textit{I felt that I didn't know what to say once I started voice input, and I couldn't properly convey my thoughts.}''
With the keyboards, our participants were relaxed and could enter calmly: ``\textit{Also, when it came to speech, I was too nervous to say what I was thinking, so texting was easier for me since I could slowly look back and express my thoughts.}'' (PA2).

\subsection{Stress-relieving Effect \& Content of Expression}

We analyzed quantitative data on stress relief and expressive writing. Both \textit{Keyboard} and \textit{Voice} conditions promptly reduced stress, with participants experiencing greater relief with \textit{Keyboard}. Expression content showed minimal change.

\paragraph{Immediate Stress Change}
\begin{figure*}[tbp]
    \label{fig:Study2-CatharticEffect}
    \centering
    \begin{minipage}[b]{0.4\linewidth}
        \centering
        \includegraphics[keepaspectratio, width = 0.6\linewidth]{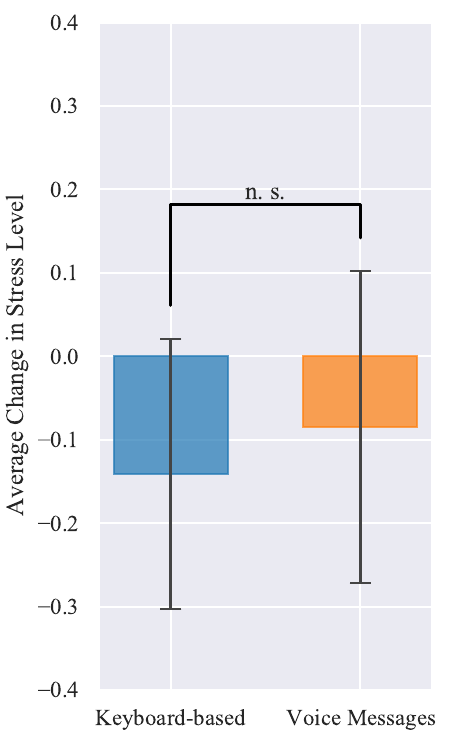}
\subcaption{The average differences in stress level before and after emotional expression.}
        \label{fig:change}
    \end{minipage}
    \hspace{0.1cm}
    \begin{minipage}[b]{0.5\linewidth}
        \centering
        \includegraphics[keepaspectratio, width=0.48\linewidth]{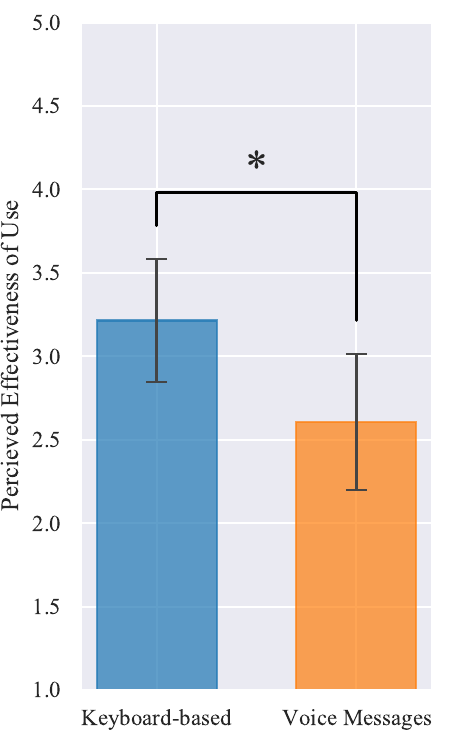}
        \subcaption{The average perceived effectiveness in reducing stress collected in the post-experiment questionnaire.}
        \label{fig:exit-effectiveness}
    \end{minipage}
    \caption{Results related to perceived cathartic effects immediately after emotional expression and overall effectiveness in reducing stress in Study 1.
    All error bars indicate 95\% confidence interval. (*: $p < .05$, **: $p < .01$, ***: $p < .001$)}
\end{figure*}

The average values of \textit{immediate stress change} were -0.14 ($SD$ = 0.44) and -0.085 ($SD$ = 0.50) under the \textit{Keyboard} and \textit{Voice} conditions, respectively (Figure~\ref{fig:change}).
Note that we gained \textit{immediate stress change} by calculating each participant's average ratings on each feedback.
We ran a Mann-Whitney's U test to evaluate the difference in the cathartic effect, but we could not find a significant effect ($U$ = 27087, $Z$ = -0.67, $p$ = .47, $r$ = 0.03).

\paragraph{Post-experiment Cathartic Effect}
Figure~\ref{fig:exit-effectiveness} presents the average cathartic effect ratings: \textit{Keyboard} was 3.21 ($SD$ = 0.99) and \textit{Voice} was 2.61 ($SD$ = 1.10). The \textit{Keyboard} text entry was rated significantly higher than the \textit{Voice} message, as shown by the Wilcoxon signed-rank test ($W$ = 1, $Z$ = -3.85, $p$ < .05, $r$ = 0.73). A significant positive correlation existed between the cathartic effect and perceived ease of use for both \textit{Keyboard} ($Spearman's\:r(26)$ = .52, $p$ < .01) and \textit{Voice} ($Spearman's\:r(26)$ = .71, $p$ < .001).

\paragraph{Length}
The average length of emotional expression data was 65.2 characters ($SD$ = 43.8) under the \textit{Keyboard} condition and 63.6 ($SD$ = 38.4) under the \textit{Voice} condition in Japanese~\footnote{Each data would correspond to roughly 25--32 words in English on average, for one English word experimentally corresponds to 2--2.5 characters in Japanese in terms of word/character counts.}.
Our Wilcoxon signed-rank test did not reveal a significant difference ($W$ = 1, $Z$ = 0.73, $p$ = 0.48, $r$ = 0.14).

\paragraph{Emotional Words}
The use of words related to emotions also had no significant difference between the two conditions.
Our Wilcoxon signed-rank test did not find a significant effect of input modality on the appearance frequency of positive nor negative emotion words (positive emotion: $W$ = 1, $Z$ = -1.50, $p$ = .50, $r$ = 0.29; negative emotion: $W$ = 1, $Z$ = -0.096, $p$ = .93, $r$ = 0.018).
Input modality did not affect the content of expressive writing texts.

These four quantitative metrics led us a conclusion that the two input modalities did not exhibit clear differences with respect to stress-relieving effect as well as the content in expressive writing.

\subsection{Summary}

Our first study confirmed that the keyboard-based text entry provided better usability and was preferred more because of its privacy friendliness and reviewing process.
On the other hand, the two input modalities did not show clear differences in immediate stress relief, and pre- and post-experiment stress levels showed no significant difference.
We, therefore, kept working on improving the interface design for mobile expressive writing.
We moved our attention from input to output and examined how visual feedback could impact perceived cathartic effects.

\section{Study 2: Examination of Effects of Visual Feedback}

As mentioned earlier, to further explore design possibilities for enhancing the effects of mobile expressive writing, we conducted an experiment related to RQ2, examining the impact of visual feedback that materializes emotional catharsis.
This section describes the design of our second user study.
We obtained a separate approval for the study protocol described in this section form the Institutional Review Board of the first author's institution.

\subsection{Experiment Design}

We used the same system architecture, adding visual feedback for emotional writing. Based on the first study, we chose a keyboard-based text entry method. Additionally, we allowed participants to share positive episodes when they did not have stressful moments although this data was not used in our analysis.

\subsubsection{Independent Variables: Visual Feedback Conditions}

We prepared three visual feedback conditions:  \textit{None}, \textit{Passive}, and \textit{Interactive}.
The \textit{None} condition offered no visual feedback, which is equivalent to the configuration used in Study 1, and serves as a baseline.
For the visual feedback conditions, we grounded our design choices in prior research on materialized emotional disposal as referred to in section~\ref{subsec:rw-feedback}.

\begin{figure*}[tbp]
    \centering
    \includegraphics[width=1.0\linewidth]{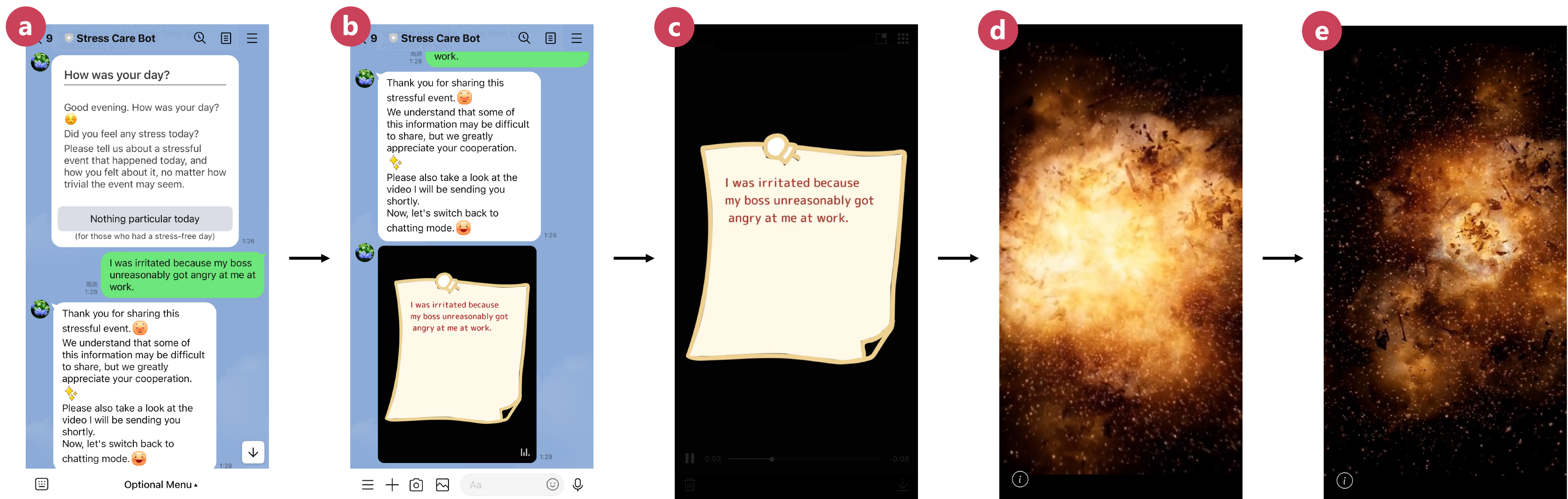}
    \caption{The \textit{Passive} feedback walkthrough: (a) Users perform emotional expression and receive a thank-you message. (b) A generated video follows. (c, d, e) Clicking reveals participant text, which then visually explodes.
    }
    \label{fig:study2-passive}
\end{figure*}

\begin{figure*}[tbp]
    \centering
    \includegraphics[width=1.0\linewidth]{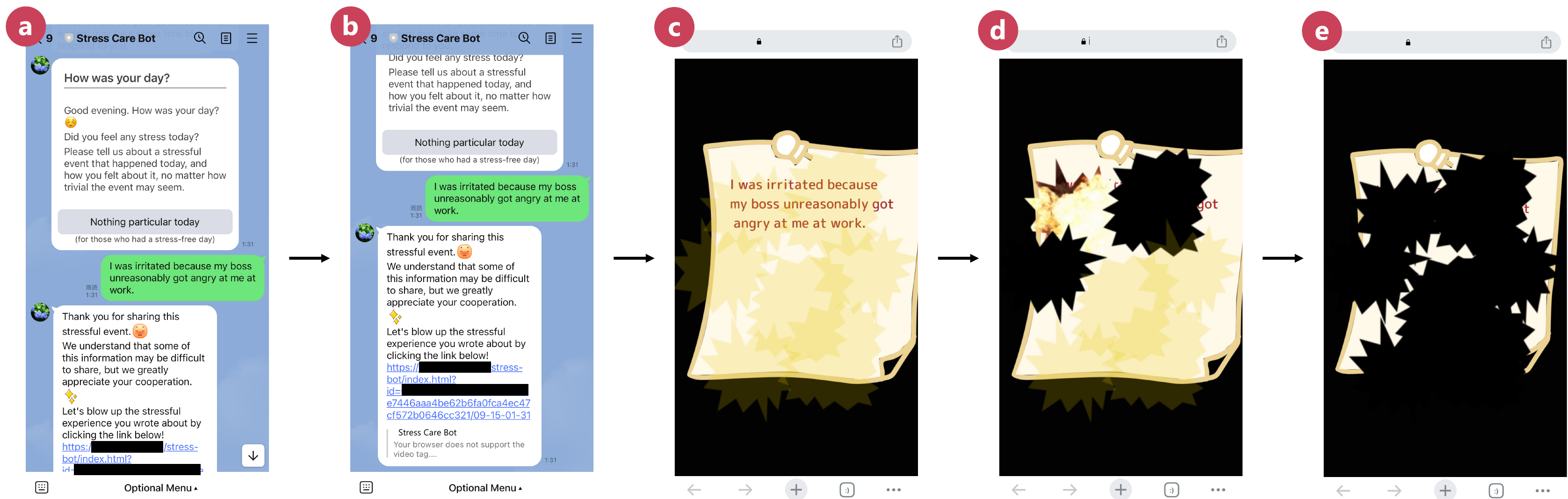}
    \caption{The \textit{Interactive} feedback walkthrough: (a) Users express emotions on request. (b) The system thanks them with a website link. (c) Clicking shows their note in a browser. (d) Tapping the note triggers an explosion effect. (e) After seven taps, the process concludes.
    }
    \label{fig:study2-interactive}
\end{figure*}

\paragraph{\textit{Passive} Feedback Condition}

In this condition, participants received unembodied, non-interactive visual feedback. When they submitted stress-related texts, the system showed a short video with their text on a note that exploded.
Figure~\ref{fig:study2-passive} details this process. We used an explosion effect similar to previous VR studies~\cite{grieger2021trash, bjorling2019thought}.
This design assumes that conveying the meaning of disposal alone can be effective without direct physical engagement~\cite{Brinol2013}, and that passive interventions are particularly suitable for high-stress contexts requiring minimal effort~\cite{zhao2023affective, DigitalWorkplace}.

\paragraph{\textit{Interactive} Feedback Condition}

In this condition, participants were offered interactive visual feedback. After submitting their stress-related text, participants could tap to explode a note with their text (Figure~\ref{fig:study2-interactive} shows the walkthrough).
This design draws on grounded cognition theory~\cite{barsalou2008grounded}, suggesting that physical engagement strengthens the perceived meaning of actions and supports emotional de-fusion~\cite{lee2021grounded, hatvany2018becoming}. It also promotes a sense of agency~\cite{moore2016sense}, potentially enhancing cathartic effects.

\vspace{0.5\baselineskip}By comparing these conditions, we aimed to examine how varying levels of user engagement in emotional disposal influence the psychological outcomes in mobile expressive writing.

\subsubsection{Dependent Variables for Stress-relieving Effect}

Similarly to our first study, we collected \textit{immediate stress change} and \textit{post-experiment cathartic effect}.
Additionally, we gathered the following quantitative data to assess the effects on mental health conditions beyond subjective impressions.

\paragraph{Long-term Stress Change}
We used the Perceived Stress Scale (PSS)~\cite{cohen1994perceived} to measure changes in participants' stress levels over time. This tool has 14 questions on a five-point Likert scale (0: never to 4: very often), assessing the past week's experiences. Weekly scores were calculated by subtracting the initial score from the later score related to the condition experienced that week.

\subsubsection{Dependent Variables for Usability and Preference}

We collected \textit{ease of use}, \textit{mental workload}, and \textit{preference} using the same questions as Study 1.
We also collected \textit{effort}, by asking participants to rate the statement, ``This feedback requires effort,'' on a five-point Likert scale (1: ``Not at all'' -- 5: ``Strongly agree''), based on prior research comparing usability of digital coping methods~\cite{DigitalWorkplace}.

\subsection{Procedure}

\label{subsec:Procedure2}
The study lasted for 23 days (three full weeks for data collection and two days for the onboarding and post-experimental stages) to accommodate the three conditions (\textit{None}, \textit{Passive}, and \textit{Interactive}).

The process in the onboarding and post-experimental sessions was the same as in Study 1.
In this study, we explained our participants the objective of our study to examine the effects of three feedback designs at this time.
The pre-experiment questionnaire included the questions from PSS for long-term stress change.
The onboarding required approximately 30 minutes to complete on the first day.
We created six groups of participants with considering the balance of gender, age, and occupations to fully counter-balance the order of the presentations of the three feedback conditions.

Participants are under each of the visual feedback conditions for each week of the experiment (Day 2--8, Day 9--15, and Day 16--22).
On the mornings of days 9 and 16, we asked participants to fill out PSS through Google Forms to track their long-term stress level trends.

We conducted post-experiment questionnaires (Table \ref{tab:study2-questionnaire} in Appendix \ref{appendix:study2-questionnaire}) on the final day, taking approximately 30 minutes.
They included PSS questions and items on \textit{post-experiment cathartic effect}, \textit{effort}, \textit{ease of use}, \textit{mental workload} of each feedback condition, and the \textit{preference} of feedback conditions. We gathered qualitative data through open-ended questions on their views about the pros and cons of each feedback condition, reasons for their preferences, preferred usage times, and desired feedback designs, aiming to better understand the quantitative results.

We offered all participants approximately 21 USD in the local currency for the completion of the study.
In consideration of the longer study duration and increased expected workload compared to Study 1, we provided a higher level of compensation..
We performed data analysis in the same way as we did in our first study.

The ethical considerations for Study 2 followed the same procedures described in Study 1.
The only difference was that Study 2 did not include deception or a subsequent debriefing.

\subsection{Participants}\label{subsec:Participants2}
We used the same crowdsourcing service as Study 1.
We also followed the same procedure to choose participants.
As a result, we recruited 36 participants (18 women, 18 men; PB1 -- 36) in their 20s to 60s (Table \ref{tab:demographics2}).
All participants self-claimed that they did not know of the existence of our first study.
Additionally, we verified that none of the participants had usernames matching those from Study 1.

\begin{table*}[t]
    \centering
    \caption{Demographics information of participants in Study 2.}
    \label{tab:demographics2}
    \begin{tabular}{|c||c|c|}
    \hline
    Groups&Gender&Age Ranges \\ \hline\hline
    Group 1 (\textit{None} $\rightarrow$ \textit{Passive} $\rightarrow$ \textit{Interactive})&
    F: 3, M: 3 &
    20s: 0, 30s: 3, 40s: 2, 50s: 1, 60s: 0\\ \hline
    Group 2 (\textit{None} $\rightarrow$ \textit{Interactive} $\rightarrow$ \textit{Passive})&
    F: 3, M: 3 &
    20s: 1, 30s: 2, 40s: 2, 50s: 1, 60s: 0\\ \hline
    Group 3 (\textit{Passive} $\rightarrow$ \textit{None} $\rightarrow$ \textit{Interactive})&
    F: 3, M: 3 &
    20s: 0, 30s: 3, 40s: 2, 50s: 1, 60s: 0\\ \hline
    Group 4 (\textit{Passive} $\rightarrow$ \textit{Interactive} $\rightarrow$ \textit{None})&
    F: 3, M: 3 &
    20s: 1, 30s: 2, 40s: 2, 50s: 1, 60s: 0\\ \hline
    Group 5 (\textit{Interactive} $\rightarrow$ \textit{None} $\rightarrow$ \textit{Passive})&
    F: 3, M: 3 &
    20s: 1, 30s: 2, 40s: 2, 50s: 0, 60s: 1\\ \hline
    Group 6 (\textit{Interactive} $\rightarrow$ \textit{Passive} $\rightarrow$ \textit{None})&
    F: 3, M: 3 &
    20s: 1, 30s: 2, 40s: 2, 50s: 1, 60s: 0\\ \hline
    \end{tabular}
\end{table*}

\section{Study 2 Results}\label{sec:results}
In this section, we present the results of the second study, covering stress reduction and usability through both quantitative and qualitative analyses.
We collected 550 expressive writing entries with immediate stress change data, and the response rate was 99.4\%.
Among the 550 expressive writing entries, 186, 183, and 181 were under the \textit{None}, \textit{Passive}, and \textit{Interactive} conditions, respectively.
The total length of responses across 16 open-ended questions averaged approximately 557 characters per participant, ranging from 181 to 1882 characters, with a response rate of 100\%.

\subsection{Quantitative Results for Stress-relieving Effect}

\begin{figure*}[tbp]
    \label{fig:result2-PerceivedEffectiveness}
    \centering
    \begin{minipage}[t]{0.47\linewidth}
        \centering
        \includegraphics[keepaspectratio, width=0.85\linewidth]{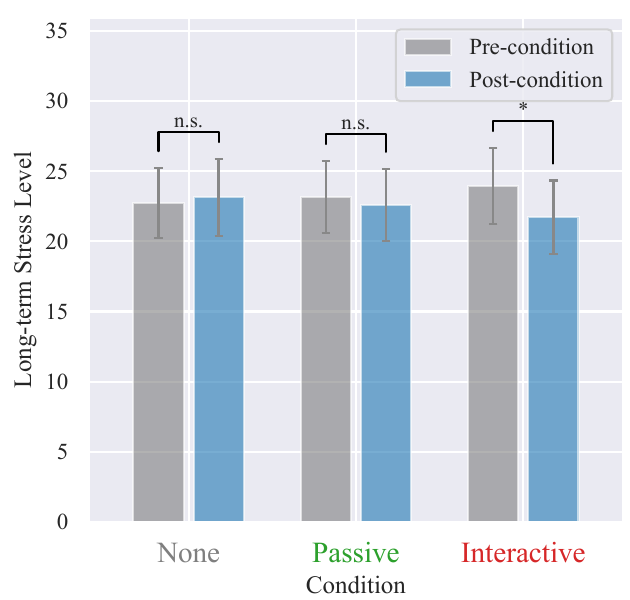}
        \subcaption{The average score of long-term stress level collected pre- and post-condition.}
        \label{fig:result2-pss_each}
    \end{minipage}
    \hspace{0.02\linewidth}
    \begin{minipage}[t]{0.47\linewidth}
        \centering
        \includegraphics[keepaspectratio, width=0.6\linewidth]{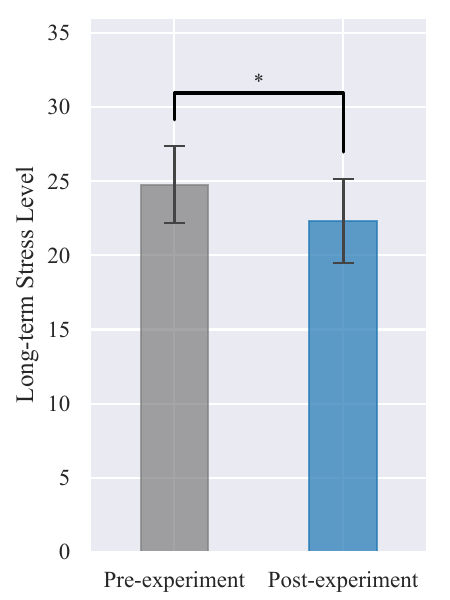}
        \subcaption{The average score of long-term stress level collected pre- and post-experiment.}
        \label{fig:result2-pss_overall}
    \end{minipage}
    \\
    \begin{minipage}[t]{0.47\linewidth}
        \centering
        \includegraphics[keepaspectratio, width = 0.6\linewidth]{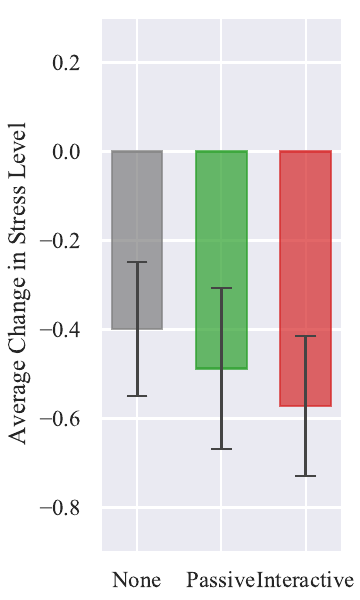}
        \subcaption{The average differences in stress level before and after expression.}
        \label{fig:result2-change_overall}
    \end{minipage}
    \hspace{0.02\linewidth}
    \begin{minipage}[t]{0.47\linewidth}
        \centering
        \includegraphics[keepaspectratio, width=0.6\linewidth]{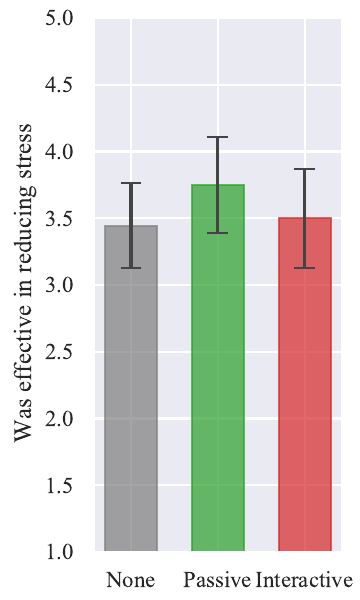}
        \subcaption{The average rating of the perceived effectiveness in reducing stress collected in the post-experiment questionnaire.}
        \label{fig:result2-effectiveness}
    \end{minipage}
    \caption{Results related to stress-relieving effect in Study 2.
    All error bars indicate 95\% confidence interval. (*: $p < .05$, **: $p < .01$, ***: $p < .001$)}
\end{figure*}

\paragraph{Long-term Stress Change}
The average difference between pre- and post-experimental PSS scores were 0.39 ($SD$ = 5.23), -0.58 ($SD$ = 5.03), and -2.22 ($SD$ = 7.06) in the \textit{None}, \textit{Passive}, \textit{Interactive} conditions, respectively (Figure \ref{fig:result2-pss_each}).
Our Friedman test did not reveal any significant difference ($\chi^2(2)$ = 1.22, $p$ = .54, $r$ = .117).
For the \textit{Interactive} condition, our Wilcoxon signed-rank test found a significant difference between pre- and post-experimental phases ($W$ = 1, $Z$ = 1.05, $p$ < .05, $r$ = .175).
For conditions \textit{None} and \textit{Passive}, the Wilcoxon signed-rank test showed no significant differences pre- and post-test (\textit{None}: $W$ = 1, $Z$ = 0.267, $p$ = .49, $r$ = 0.0445; \textit{Passive}: $W$ = 1, $Z$ = 0.0393, $p$ = .25, $r$ = 0.00655).

However, using our system for three weeks resulted in a significant stress reduction.
The average PSS scores in the pre- and post-experimental phases were 24.75 ($SD$ = 7.93) and 22.33 ($SD$ = 8.66), respectively (Figure \ref{fig:result2-pss_overall}).
Our Wilcoxon signed-rank test revealed a significant difference the pre- and post-experimental phases ($W$ = 1, $Z$ = 0.448, $p$ < .05, $r$ = 0.0746).
Experiencing mobile expressive writing through the system for three weeks thus significantly reduced the stress.

\paragraph{Immediate Stress Change}\label{subsubsec:results2-cathartic-immediate}

The average scores of \textit{immediate stress change} were -0.40 ($SD$ = 0.46), -0.49 ($SD$ = 0.55), and -0.57 ($SD$ = 0.48) under \textit{None}, \textit{Passive}, and \textit{Interactive} conditions, respectively (Figure~\ref{fig:result2-change_overall}). We calculated \textit{immediate stress change} from participants' average ratings on each feedback. The Friedman test showed no significant difference ($\chi^2(2)$ = 2.64, $p$ = .27, $r$ = 0.25).

\paragraph{Post-experiment Cathartic Effect}\label{subsubsec:results2-cathartic-retrospect}
Figure~\ref{fig:result2-effectiveness} shows the average ratings of all the three conditions: 3.44 ($SD$ = 0.97), 3.75 ($SD$ = 1.11), and 3.5 ($SD$ = 1.13) under the \textit{None}, \textit{Passive}, and \textit{Interactive} conditions respectively.
Our Friedman test did not reveal a significant difference ($\chi^2(2)$ = 2.14, $p$ = .34, $r$ = 0.21).

\subsection{Qualitative Results for Stress-relieving Effect}

From our thematic analysis, we revealed that each condition provided the cathartic effect differently.
The variety in the existence of \textit{visual feedback} and \textit{agency} created different outcomes.

\subsubsection{Two Visual Feedback Designed Provided a Sense of Catharsis but in Different Ways.}

Participants felt the cathartic effects in \textit{Passive} and \textit{Interactive} conditions, yet differences arose in agency of elimination. In \textit{Interactive} condition, exploding letters themselves provided relief for seven participants: ``\textit{I myself had exploded the stress, and it felt refreshing.}'' (PB32). Interactivity influenced stress relief agency, with lack of it being a downside in \textit{Passive} condition—they felt \textit{``just like observing''} (PB28) and out of control. Control over visual stress disappearance enhanced the cathartic effect.

Participants experienced cathartic benefits from the \textit{Passive} condition by distancing from stress. Laughter, noted by five participants, was an advantage of the \textit{Passive} condition: ``\textit{It was good that it was simply amusing and made me laugh. I thought it was especially good when I was feeling stressed out and bothered.}'' (PB11). The clear visualization of stress-related emotion release also refreshed or calmed them: ``\textit{Seeing the stress words explode and disappear made me feel a bit amusing in a `deserve it!' kind of way, and I appreciated that it made me feel calm.}'' (PB30).

\subsubsection{No Feedback Still Offered Reflection Opportunities.}

Our qualitative results did not find explicit benefits of catharic effects in the \textit{None} condition.
PB30, for example, pointed out its disadvantage of not offering an explicit way to vent: ``\textit{When there is too much stress, thinking about which ones to list or write all of them, or remembering that stress, causes extra stress. I feel like it’s a waste of time to dwell on it, remember it, and go back and forth over it. That’s what I find problematic.}''

However, our participants saw the value of this condition outside the cathartic effect.
Participants noted that the system actively listened to their expressions and appreciated the immediate responses. The pre-defined reply message\footnote{The message used in the second user study was ``Thank you for sharing this stressful event. We understand that some of this information may be difficult to share, but we greatly appreciate your cooperation.''}, used in all conditions, provided them feedback and confirmation of being heard.
``\textit{It was beneficial to be able to directly describe the accumulated stress of the day to the chatbot.
The immediate response from the chatbot made me feel like it was actively listening to my thoughts.}'' (PB26)

Participants also found benefit in organizing their thoughts. Seven noted they could recognize and contemplate stress: ``\textit{I was able to reflect on the events that caused stress and think about them logically. It was beneficial to calmly reconsider the origins of feelings of aversion and anger towards others.}'' (PB11). This implies that the absence of visual feedback may encourage reflection, enhancing expressive writing's benefits in a different manner from the conditions with visual feedback.

\subsection{Quantitative Results for Usability \& Preference}\label{subsubsec:results2-easeofuse}
With the qualitative analysis of usability and preference metrics, we found that the \textit{Interactive} condition required more effort and mental workload than the \textit{Passive} condition did.
However, the provision of visual feedback did not compromise the usability of the system, and the participants exhibited a neutral preference for the conditions.

\begin{figure*}[tbp]
    \centering
    \begin{minipage}[t]{0.32\linewidth}
      \centering
      \includegraphics[keepaspectratio, width = 0.77\linewidth]{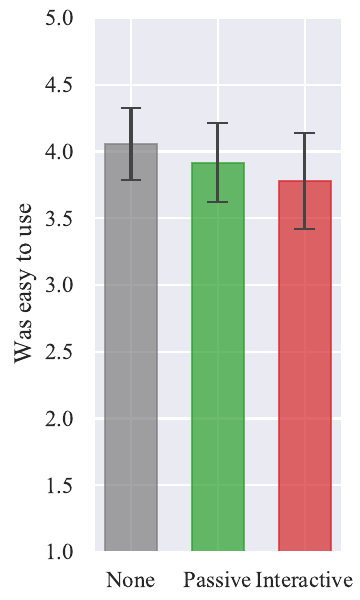}
      \subcaption{The average perceived ease of use.}
      \label{fig:result2-easeofuse}
    \end{minipage}
    \hspace{0.01\linewidth}
    \begin{minipage}[t]{0.32\linewidth}
      \centering
      \includegraphics[keepaspectratio, width = 0.77\linewidth]{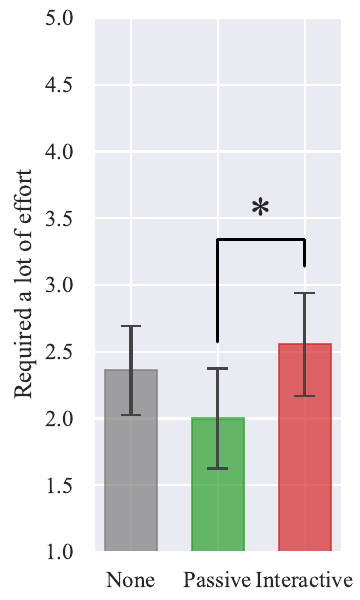}
      \subcaption{The perceived effort required.}
      \label{fig:result2-effort}
    \end{minipage}
    \hspace{0.01\linewidth}
    \begin{minipage}[t]{0.32\linewidth}
      \centering
      \includegraphics[keepaspectratio, width = 0.77\linewidth]{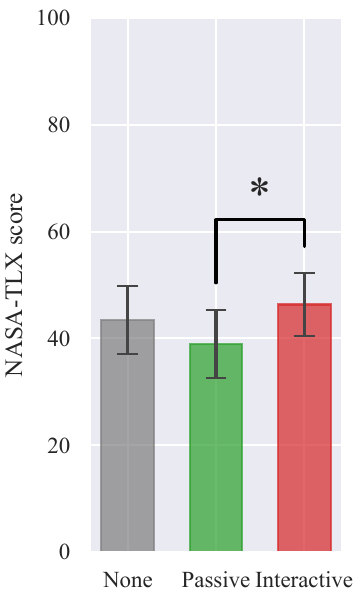}
    \subcaption{The average mental workload scores.}
    \label{fig:result2-nasatlx}
    \end{minipage}
  \caption{Results related to the usability of the three feedback conditions. All were collected through the post-experiment questionnaire in Study 2.
  All error bars indicate 95\% confidence interval. (*: $p < .05$, **: $p < .01$, ***: $p < .001$)}
  \label{fig:result2-Ease_and_Load}
\end{figure*}

\paragraph{Ease of Use}
The average ratings of \textit{ease of use} were 4.06 ($SD$ = 0.83), 3.92 ($SD$ = 0.91), and 3.78 ($SD$ = 1.10) under the \textit{None}, \textit{Passive}, and \textit{Interactive} conditions respectively (Figure~\ref{fig:result2-easeofuse}).
Our Friedman test did not show a significant difference ($\chi^2(2)$ = 0.42, $p$ = .81, $r$ = 0.04).

\paragraph{Effort}
The average ratings of \textit{effort} were 2.36 ($SD$ = 1.02), 2.00 ($SD$ = 1.15), and 2.56 ($SD$ = 1.18) under the \textit{None}, \textit{Passive}, and \textit{Interactive} conditions respectively (Figure~\ref{fig:result2-effort}), confirming a significant difference with our Friedman test ($\chi^2(2)$ = 10.18, $p$ < .01, $r$ = 0.98).
The post-hoc Nemenyi test confirmed significant differences between \textit{Passive} and \textit{Interactive} ($p$ < .05).

\paragraph{Mental Workload}
The average \textit{mental workload} scores were 43.4 ($SD$ = 19.5), 39.0 ($SD$ = 19.67), and 46.35 ($SD$ = 17.95) under the \textit{None}, \textit{Passive}, and \textit{Interactive} conditions respectively (Figure~\ref{fig:result2-nasatlx}).
Our Friedman test revealed a significant difference ($\chi^2(2)$ = 7.90, $p$ < .05, $r$ = 0.760).
The post-hoc Nemenyi test confirmed significant differences between \textit{Passive} and \textit{Interactive} ($p$ < .05).

\paragraph{Preference}
In the post-experiment questionnaire, 9, 14, and 13 participants mentioned the \textit{None}, \textit{Passive}, and \textit{Interactive} conditions as the most preferred feedback, respectively.
Our chi-square test did not reveal that participants preferred specific conditions ($\chi^2(2)$ = 1.17, $p$ = 0.56).

\subsection{Qualitative Results for Preferences}

\subsubsection{Scenario-dependent Preferences}\label{subsec:results2-usescenarios}
Our qualitative results uncovered different rationales for preferences of each condition.
The \textbf{\textit{None} condition} was found useful when they simply wanted someone else to hear their complaints.
Our participants also commented that it could help them  reflect on their feelings through writing or leaving records for future reflections: ``\textit{I want to use it (the \textit{None} condition) when I can't talk to others about something, but it's hard for me to keep it to myself, and I want to be heard, even if it's by a robot.}'' (PB29) 

Our participants expressed that they would use the \textbf{\textit{Passive} condition} when the stress originated from others.
Other participants also explicitly mentioned that they would rather avoid using the \textit{Passive} design when the cause of their stress was associated with themselves: ``\textit{I want to use it (the \textit{Passive} condition) when I feel stressed about others. I would not use it for stress within myself because I don't feel it is very effective.}'' (PB13)

The \textbf{\textit{Interactive} condition} was deemed useful when participants felt intense stress.
However, our participants felt that they would not use it when they were exhausted: ``\textit{I wanted to use the \textit{Interactive} condition when I was feeling a bit strongly stressed. There are quite a few times when expressions alone are not enough to completely relieve stress, so I felt that this condition would help me relieve even the intense stress.}'' (PB33)
This impression is in line with our quantitative results on the perceived effort, where the \textit{Interactive} condition was rated as significantly more effortful than the \textit{Passive} condition (Section~\ref{subsubsec:results2-easeofuse}).

Our qualitative results reveal scenarios for the three visual conditions not shown in the quantitative data. Participants found the \textit{None} condition useful for mental cool-down and relaxation. Comments highlighted that both the \textit{Passive} and \textit{Interactive} conditions served better for stress from others.
The \textit{Interactive} design helped to relieve intense stress though subjective workload was higher in the other designs.

\subsubsection{Other Potential Feedback Designs}\label{subsec:results-suggestions}
Our participants offered various ideas on other possible visual feedback designs.
They are categorized into three design principles: \textit{immediate destruction}; \textit{gradual disappearance}; and \textit{detachment}.
While the \textit{Passive} and \textit{Interactive} conditions represent \textit{immediate destruction}, our participants suggested other designs, such as braking blocks and letting virtual characters eat text.
\textit{Gradual disappearance} is visual effects where expressed negative experience would slowly disappear, such as a fade-away effect of the written text. 
Our participants commented that such effect may offer more calm feeling for stress relief: ``\textit{A video where the text gradually fades out feels calming, making me want to use it. I thought the explosion effect was too exciting for stress relief and not very effective.}'' (PB30)
\textit{Detachment} is intended to allow users to simply leave their negative experience outside themselves.
This design principle matches with the \textit{None} condition, which may promote reflection.
Our participants suggested a calm visual feedback that is not directly related to the written experience, such as flowers blooming up and animals playing around: ``\textit{I think something soothing or relaxing, like animals, would be good.}'' (PB8)

\subsection{Summary}

In our second study, participants reported a significant reduction in stress levels after completing the experimental sessions.
However, we found no significant differences in cathartic effects among the visual feedback conditions.
Qualitative analysis further revealed distinct ways each visual feedback influenced users' perceptions and emotional experiences.
In the next section, we discuss these findings in relation to our original research questions and to relevant literature.

\section{Discussion}

\subsection{Advantages of Keyboard-based Text Entry in Mobile Expressive Writing}

We conducted Study 1 to answer RQ1, which highlighted the unique benefits of keyboard-based text entry over voice input.
Aligned with previous psychology research~\cite{murray1994emotional}, our results showed that keyboard-based and voice inputs provided similar cathartic effects. However, keyboard text entry slightly outperformed voice input in usability and preference. This challenges past findings of speech input's superiority in different contexts~\cite{kim2018use, luo2022promoting, rzepka2022voice, kocielnik2018designing}.
Qualitative insights suggest that the affordances of keyboard entry helped ensure confidentiality~\cite{easwara2015privacy} but also better supported the cognitive restructuring process~\cite{pennebaker2007expressive} central to expressive writing.

Participants favored keyboard-based text entry
for its persistence and editability~\cite{treem2013social}, enhancing two key psychological mechanisms
of expressive writing: \textit{exposure}~\cite{sloan2004taking} and \textit{cognitive processing}~\cite{pennebaker2007expressive}.
Typing enables the externalization of one's stress in a persistent, reviewable format~\cite{clark1991grounding}, functioning as an exposure to past stressors --- an important process in expressive writing~\cite{sloan2004taking}.
Moreover, the ability to revise enabled participants to reconstruct their narratives~\cite{pennebaker2007expressive}.
From the perspective of cognitive processing theory, expressive writing involves a dynamic process of planning, translating, and reviewing, rather than a simple linear flow of thoughts~\cite{flower1981cognitive}.
In this regard, typed input enabled users to pause, reflect, and revise their expressions~\cite{daiute1986physical}, thereby supporting narrative reorganization.
In contrast, voice input encouraged a more linear, less editable form of narration~\cite{schalkwyk2010your, lai2006speech}, which made it harder for participants to revise or redirect their thoughts once expressed.
Previous research has similarly shown that typed input on mobile devices offers superior editing capabilities compared to speech input~\cite{cherubini2009text, luo2022notewordy}, which may facilitate deeper cognitive processing through iterative reflection and reinterpretation~\cite{pennebaker2007expressive}.

On the other hand, Murray et al.~\cite{murray1989emotional} and Esterling et al.~\cite{esterling1994emotional} reported that expressing with voice into a tape recorder included more content reflecting cognitive changes and negative emotions.
While using the same modality to express stressful experiences, our study did not show similar outcomes.
This comparison proposes that chat interface may bring \textit{audience impairment effects}~\cite{lord1987effects}.
Some of our participants felt nervous when using voice input, possibly because they were conscious of the presence of the researcher as the recipient.
The audience effect is known to influence behavior even when the audience is merely imagined~\cite{higgs1971effects}, and its impact tends to be stronger when the imagined audience is someone with whom the individual has low involvement~\cite{axsom1987audience}.
Taken together, these findings suggest that in a chat interface, sending a voice message may evoke a strong sense of being monitored while recording.
This perceived observation could create pressure that disrupts cognitive processing.

Building on these findings, designing the voice-based mobile interface for expressive writing requires addressing both privacy concerns and the \textit{audience impairment effect}.
To mitigate this, systems should foster a sense of psychological safety.
One approach is to reduce an audience's perceived presence, for example, by making it clear that recordings are stored only locally and not automatically transmitted or by explicitly separating the recording and sending processes to give users more control.
Alternatively, designers could embrace the presence of an audience by providing responsive voice feedback, such as affirmations or empathetic cues, similar to how text-based chatbots build affirmative relationships through empathic responses~\cite{fitzpatrick2017delivering}.
These strategies highlight the importance of carefully shaping the social dynamics users perceive during voice-based emotional expression.

\subsection{Feedback Design Preferences Affected by Multiple Factors}

The results of Study 2 to answer RQ2 revealed distinct characteristics and benefits for all three feedback conditions, along with their possible use scenarios.
Previous work has shown how post-disclosure interaction helps alleviate negative feelings toward certain topics, e.g., by disposing of the paper on which envious feelings were disclosed, compared to a baseline of doing nothing~\cite{Brinol2013, grieger2021trash, soesilo2021no}.
We did not, however, observe a similar effect in our study as there were no significant differences between the conditions concerning stress release.
Our results suggest that digital tools like what we developed in this study, while effective, may offer different effects from what is observed between the physical act of discarding a paper into a bin and taking no action, encouraging further research.

By allowing participants to express stress in any context in their lives, our study covered a diverse set of conditions and sources of stress. 
In our case, particularly whether the stress was self-caused or had something to do with other people, i.e., the source, was one of the determining factors to the perceived appropriateness of feedback designs.
Related to this effect of the source, Bri{\~n}ol et al. found that negative emotions related to participants themselves persisted even after tearing up their written thoughts, while negative emotions about others disappeared~\cite{Brinol2013}.
Our findings help confirm the effect of the source with a broader set of stressful moments people may face in their daily lives, expanding the existing understanding of the effect of expressive writing.

Our study also found that the \textit{Passive} condition, where the feedback did not provide a sense of agency, induced laughter.
The incongruity between expectation and outcomes can generate laughter~\cite{rothbart1973laughter}.
The unexpected, exaggerated explosion provoked laughter, and the laughter overcame the negative emotion~\cite{keltner1997study}.
The combination of both objectified emotion and the act of laughter may augment the cathartic effect.

The \textit{Interactive} condition, where users tapped to trigger the visual disappearance, was perceived as more effective in relieving stress.
Participants attributed this effect to their ability to actively control the experience.
This sense of agency~\cite{moore2016sense, alhasani2022systematic} appeared to amplify the cathartic release of \textit{inhibited} emotions beyond what the writing process alone could offer~\cite{pennebaker1986confronting}.
Providing users with opportunities to act at their own pace can enhance their sense of agency and personal involvement, which are known to support emotional regulation.
However, our findings also indicate that increased interactivity may introduce additional cognitive or physical workload.
To address this trade-off, it may be beneficial to design interventions that offer adjustable levels of agency.
For instance, system could allow users to choose between passive and interactive modes, or dynamically adapt the level of interactivity based on contextual factors or intensity of stress.
Such flexibility may improve both user engagement and the overall effectiveness of the system.

Participants also expressed that the \textit{None} design was also valuable for reflecting on their emotions.
One possible explanation for this effect is that this condition may have offered an environment of active listening.
Active listening, which is one of the key skills in counseling~\cite{hutchby2005active}, supports the speakers to be more emotionally mature and open to their experiences~\cite{rogers1957activelistening}.
Our system sent participants notifications after they submitted their disclosure to confirm its reception, but it may have served as a way to demonstrate listening. 

Based on our results, stress-coping interventions should consider the characteristics of stress to improve their acceptability.
A growing body of research has developed stress-coping support systems that deliver just-in-time adaptive interventions (JITAI)~\cite{nahum2018just, han2022stressbal}.
Recent adaptive systems often utilize biometric data (e.g., heart rate collected from smartwatches~\cite{han2022stressbal, wakschlag2021moving}) or self-reported data (e.g., stress levels collected via ecological momentary assessment, or EMA~\cite{juarascio2021clinician, presseller2022self}) as tailoring variables to determine the optimal timing of interventions.
By scaling the level of stress based on these data, these systems have successfully delivered interventions when users are most vulnerable, supporting more informed decision-making.
However, an equally important factor—receptivity—has largely been overlooked~\cite{van2025beyond}.
In light of our findings, we suggest that effective intervention design should consider not only the intensity of stress but also the orientation of stress when determining the nature and timing of support.
In this context, expressive writing, as an activity that involves articulating personal experiences and emotions, may play a dual role: not only as a therapeutic tool, but also as a means of sensing.
It can offer valuable insights into the sources and directions of stress, making it a promising component of mobile systems for everyday stress management and adaptive interventions.

\subsection{Research Opportunities in Mobile Expressive Writing}

Our second study indicates that visual design affects user experiences differently. Participants felt like observers in the \textit{Passive} condition and like active stress relievers in the \textit{Interactive} condition. Literature suggests personality (e.g., extrovert vs. introvert) influences preferences for visual feedback~\cite{carver2010personality}. A more extensive study on these factors can further deepen the understanding of mobile interface designs for expressive writing.
Further, when our participants received text-based feedback after disclosure, they felt like someone heard their narratives.
Here, future research could investigate how to design feedback so that users focus more on the disclosed emotions themselves rather than the feedback and its effects, which could lead to useful cognitive restructuring. 

Speech-based mobile input can provide real-time recognition and transcription~\cite{li2022recent}. 
As our study identified clear differences between text and speech inputs (Section~\ref{subsubsec:study1-results-usability}), sensitive tasks like expressive writing require careful design considerations on input methods.
Luo showed that combining keyboard and speech input for food journaling~\cite{luo2022promoting} could enhance usability and self-reflection. Future research could explore dividing tasks between keyboard and speech input for mobile expressive writing.

Smartphones can facilitate many types of rich interactions that are not available with other writing platforms. 
Auditory or haptic feedback~\cite{takami2022stressmincer} may be a possible direction to explore with expressive writing solutions
For instance, Costa et al.~\cite{costa2016emotioncheck} showed how haptic feedback of heart rate could help users regulate their anxiety.
Future systems could include deep breathing and mindfulness activities with emotional expression and examine how they could enhance cathartic effects and overall stress relief in conjunction with expressive writing tasks.
Multimodal interface and feedback designs for stress relief may thus lead to a new interdisciplinary research opportunity for HCI and psychology. 
Moreover, smartphones are excellently equipped for longitudinal data collection studies. While our focus here is on comparing the conditions (designs), a longitudinal study with a similar journaling application focusing on temporal changes would be interesting. Here, issues such as how participants perceptions evolve as a function of study progression, or whether their preferences shift over time would reveal important findings about the effects of novelty in similar deployments.

\subsection{Limitations}

We acknowledge several limitations in our work.
Our study was conducted in Japan with a culturally homogeneous sample of native Japanese speakers. This might have influenced speech input effectiveness, as verbalizing thoughts relates differently to cognitive processing in some Asian cultures compared to the West~\cite{kim2002we}. Future research should consider diverse cultural backgrounds. None of the participants frequently used speech input, which may have affected their preferences. Further, our qualitative findings indicate that keyboard input advantages extend beyond mere familiarity. 
We adapted questionnaires used in previous work to our studies, and we used the same questionnaires across the conditions to minimize the effect of question designs. Future work may further validate our results by employing different framing and styles of questions.
Despite the limitations, our work contributes two successful deployment studies on expressive writing in authentic settings, which advance the field of mobile HCI for stress management.

\section{Conclusion}

Expressing negative emotions is known to be an effective stress-coping mechanism, but challenges with respect to interface designs emerge when the practice is transitioned to technology-supported mental well-being systems. 
This work investigated how input modalities for emotional expression and visual feedback designs for post-expression interventions could affect the perceived cathartic effects and user experience.
Our study highlighted the benefits and stronger preference for keyboard-based text entry over input methods using voice messages.
It also revealed how different visual feedback designs can benefit users to accommodate different scenarios of stress-related emotional expression.
Our findings also suggest important design considerations and future research directions in mobile expressive writing, and further encourage interdisciplinary research between HCI and self-help approach in psychology.

\begin{acks}
We are grateful for the support received during the preparation of this paper.
In particular, we would like to thank Ginshi Shimojima, Chi-lan Yang, Zefan Sramek, and Shitao Fang for their valuable feedback.
We also extend our appreciation to the participants in our user study for their involvement.
This research is part of the results of Value Exchange Engineering, a joint research project between R4D, Mercari, Inc., and the RIISE.
This project was also partly supported by JSPS Bilateral (Grant Number: 120232701), Research Council of Finland (Grant Numbers 349637 and 353790), and JST ASPIRE for Top Scientists (Grant Number: JPMJAP2405).
\end{acks}

\bibliographystyle{ACM-Reference-Format}
\bibliography{ref.bib}

\clearpage
\appendix

\onecolumn

\section{Questionnaires Used in Study 1}

\label{appendix:study1-questionnaire}
The original questionnaire was in Japanese, and we translated it into English for the report in this paper.
We referred to our system as ``a chatbot'' as the interface was integrated into LINE text chat.

\begin{table*}[h]
\scriptsize
\begin{tabular}{p{5mm}p{12cm}}
\toprule
\multicolumn{2}{l}{\textbf{Pre-experiment Questionnaire}}  \\ \hline \hline
\multicolumn{2}{l}{Demographic Information} \\
 & Please provide your username on the crowdsourcing website. \\
 & Please specify your gender. \\
 & Please specify your age group. \\ \hline
\multicolumn{2}{l}{PSS questions} \\
& 14 questions from PSS about the past week. \\ \hline
 \multicolumn{2}{l}{Chatbot registration and confirmation} \\
 & Did you register the chatbot from the link above and send your username on the crowdsourcing website? (Yes) \\ \hline
\multicolumn{2}{l}{\textbf{Post-experiment Questionnaire}}  \\ \hline \hline
\multicolumn{2}{l}{PSS questions} \\
& 14 questions from PSS about the past week. \\ \hline
\multicolumn{2}{l}{Overall experience of the system}  \\
 & How easy was it to use the system? (1--5)  \\
 & Do you think providing descriptions of your stress helped in relieving stress? (1--5) \\
 & Please explain why you answered the above question as you did.   \\ \hline
\multicolumn{2}{l}{Keyboard-based text entry}  \\
 & Was describing your stress using keyboard-based text input easy to do? (1--5) \\
 & Do you believe that keyboard-based text input helped in relieving stress? (1--5) \\
 & Please share your thoughts and feedback on keyboard-based text input. \\
 & Workload (NASA-TLX) with keyboard-based text input. \\ \hline
\multicolumn{2}{l}{Voice messages} \\
 & Was describing your stress using voice messages easy to do? (1--5)  \\
 & Do you believe that voice messages helped in relieving stress? (1--5)  \\   
 & Please share your thoughts and feedback on voice messages. \\
 & Workload (NASA-TLX) with voice messages. \\ \hline
\multicolumn{2}{l}{Other questions} \\
 & Which input method do you prefer? \\
 & Why do you prefer that input method?      \\
 & Do you use voice input regularly in your daily life? (1--5) \\
 & Were you aware that the primary purpose of this survey was an experiment on input methods? (1--5)  \\
 & If you have any opinions or advice regarding stress relief through the system, please share.       \\
 & Is there anything else you would like to add?  \\
 & In the future, we may request your participation in additional surveys related to this research. May we contact you for such purposes? (Yes/No)  \\ \bottomrule
\end{tabular}
\caption{Questions used in the questionnaires in Study 1.}
\label{tab:study1-questionnaire}
\end{table*}

\newpage\onecolumn
\section{Questionnaires Used in Study 2}

\label{appendix:study2-questionnaire}
The original questionnaire was in Japanese, and we translated it into English for the report in this paper.
We referred to our system as ``a chatbot'' as the interface was integrated into LINE text chat.

\begin{table*}[h]
\scriptsize
\begin{tabular}{p{5mm}p{12cm}}
\toprule
\multicolumn{2}{l}{\textbf{Pre-experiment Questionnaire}}  \\ \hline \hline
\multicolumn{2}{l}{Demographic Information} \\
 & Please provide your username on the crowdsourcing website. \\
 & Please specify your gender. \\
 & Please specify your age group. \\ \hline
 \multicolumn{2}{l}{PSS questions} \\
 & 14 questions from PSS about the past week. \\ \hline
 \multicolumn{2}{l}{Chatbot registration and confirmation} \\
 & Did you register the chatbot from the link above and send your username on the crowdsourcing website? (Yes) \\ \hline
\multicolumn{2}{l}{\textbf{Post-experiment Questionnaire}}  \\ \hline \hline
\multicolumn{2}{l}{PSS questions} \\
 & 14 questions from PSS about the past week. \\ \hline
\multicolumn{2}{l}{Experience of None Feedback} \\ 
 & 12 questions (in Table~\ref{tab:study2-Eightquestions}) about \textit{None} feedback. \\
 & Workload during \textit{None} Feedback: NASA-TLX. \\ \hline
\multicolumn{2}{l}{Experience of \textit{Passive} Feedback} \\ 
 & 12 questions (in Table~\ref{tab:study2-Eightquestions}) about \textit{Passive} feedback. \\
 & Please share any preferences for other types of videos you would like to see. \\
 & Workload during \textit{Passive} Feedback: NASA-TLX. \\ \hline
\multicolumn{2}{l}{Experience of \textit{Interactive} Feedback} \\ 
 & 12 questions (in Table~\ref{tab:study2-Eightquestions}) about \textit{Interactive} feedback. \\
 & Please share any preferences for other types of interactions. \\
 & Workload during \textit{Interactive} Feedback: NASA-TLX. \\ \hline
\multicolumn{2}{l}{Other questions} \\
 & Where did you primarily use the chatbot? (Home/Work/Commute/Other) \\
 & How do you usually cope with stress in your daily life? (Rest, Relaxation/Exercise/Talking or meeting with family and friends/Eating favorite foods/Playing games/Drawing or playing music/Shopping or going to karaoke/Other) \\
 & Which type of feedback do you prefer to use? (None/Passive/Interactive) \\
 & Why do you prefer that type of feedback? \\
 & Do you wish to continue using this chatbot? (1: Not at all–5: Strongly Agree) \\
 & On a scale of 0–10, how likely are you to recommend this chatbot to friends or colleagues? \\
 & What other features would you like to see in addition to video replies and website usage? \\
 & Please share any opinions or advice regarding stress relief through using the chatbot. \\
 & Is there anything else you would like to add? \\
 & In the future, we may request your participation in additional surveys related to this research. May we contact you for such purposes? (Yes/No) \\
\bottomrule
\end{tabular}
\caption{Questions used in the questionnaire in Study 2. The questions about three feedback designs were presented in the order individuals experienced in the experiment.}
\label{tab:study2-questionnaire}
\end{table*}

\subsection{12 Questions about Each Feedback Design}

\label{appendix:study2-Eightquestions}

\begin{table*}[h]
\scriptsize
\begin{tabular}{p{5mm}p{12cm}}
\toprule
\multicolumn{2}{l}{Eight 5-point Likert scale questions (1: Not at all–5: Strongly Agree)} \\ 
 & \textit{This} feedback was easy to use. \\
 & \textit{This} feedback helps in reducing stress. \\
 & \textit{This} feedback makes you more inclined to provide descriptions. \\
 & \textit{This} feedback makes you more likely to use our system. \\
 & \textit{This} feedback requires effort. \\
 & You are satisfied with \textit{This} Feedback. \\
 & You intend to continue using \textit{This} Feedback. \\
 & You have complaints about \textit{This} Feedback. \\ \hline
\multicolumn{2}{l}{Four open-ended questions} \\
 & When would you like to use \textit{This} Feedback (or not use it)? \\
 & Please share the positive aspects of \textit{This} Feedback. \\
 & Please share the negative aspects of \textit{This} Feedback. \\
 & Any other opinions or feedback regarding \textit{This} Feedback. \\
\bottomrule
\end{tabular}
\caption{Eight 5-point Likert scale questions and four open-ended questions about each feedback.}
\label{tab:study2-Eightquestions}
\end{table*}

\end{document}

%% file: preamble_lab.tex
\usepackage{amsmath}
\usepackage{here}
\usepackage{tabularx}
\usepackage{url}
\usepackage{enumerate}
\usepackage[inline]{enumitem}
\usepackage[font=small,labelfont=bf,tableposition=top]{caption}
\usepackage{subcaption}
\usepackage{comment}
\usepackage{multirow}

\def\Underline{\setbox0\hbox\bgroup\let\\\endUnderline}
\def\endUnderline{\vphantom{y}\egroup\smash{\underline{\box0}}\\}
\def\|{\verb|}

\renewenvironment{quote}[1][0.04\linewidth]
  {\list{}{\leftmargin=#1\rightmargin=#1}\item\relax}{\endlist}

\usepackage{color}
\definecolor{orange}{RGB}{255,127,0}
\definecolor{limegreen}{RGB}{50, 205, 50}
\definecolor{violet}{RGB}{148,0,211}

\newif\ifCOMMENTS
\COMMENTStrue
 
\ifCOMMENTS

\else

\fi

